\documentclass[runningheads]{svmult}

\usepackage{makeidx}   % allows index generation
\usepackage{graphicx}  % standard LaTeX graphics tool
\usepackage{subeqnar}  % subnumbers individual equations
\usepackage{multicol}  % used for the two-column index
\usepackage{physprbb}  % modified textarea for proceedings,
\makeindex             % used for the subject index
\newcommand{\req}[1]{(\ref{#1})}

\newcommand{\muR}{\mu_{R}^{2}}
\newcommand{\muF}{\mu_{F}^{2}}
\newcommand{\muO}{\mu_{0}^{2}}
\newcommand{\eps}{\epsilon}

\begin{document}
\title*{BLM scale for the pion transition form factor}
\toctitle{BLM scale for the pion transition form factor}
\titlerunning{BLM scale for the pion transition form factor}
% allows abbreviation of title, if the full title is too long
% to fit in the running head
%
\author{Bla\v{z}enka Meli\'{c}\inst{1,2,3}%
%\protect\footnote{On leave of absence from the 
%Rudjer Bo\v{s}kovi\'{c} Institute,
%Zagreb, Croatia.}
\and Bene Ni\v{z}i\'{c}\inst{3}
\and Kornelija Passek\inst{3}
\protect\footnote{Talk given by K. Passek at
the 8th Adriatic Meeting, Dubrovnik, September 2001.}
}
\authorrunning{Meli\'{c}, Ni\v{z}i\'{c}, Passek}
% if there are more than two authors,
% please abbreviate author list for running head
%
%
\institute{Institut f\"{u}r Physik,
        Universit\"{a}t Mainz,
        D-55099 Mainz, Germany 
\and Institut f\"{u}r Theoretische Physik,
     Universit\"{a}t W\"{u}rzburg,\\
     D-97074 W\"{u}rzburg, Germany 
\and Theoretical Physics Division, 
        Rudjer Bo\v{s}kovi\'{c} Institute, \\
        P.O. Box 180, HR-10002 Zagreb, Croatia\\[0.15cm]
{\bf \it (September, 2001)\\[-0.3cm]}}
\maketitle              % typesets the title of the contribution

\begin{abstract}
We review 
the determination of the NLO Brodsky-Lepage-Mackenzie (BLM) 
renormalization scale for the pion transition form factor.
%A consistent calculation 
%up to $n_f$-proportional NNLO contributions to both
%the hard-scattering amplitude and the 
%perturbatively calculable part of the pion
%distribution amplitude is reviewed.
We argue that the prediction for the 
pion transition form factor is 
independent of the factorization scale 
at every order in the strong coupling constant.
%making it possible to use for calculational purposes
%the simplest choice $\muF=Q^2$.
%Assuming the pion asymptotic distribution amplitude
%and working in the $\overline{MS}$ scheme,
%we find the BLM scale to be $\muR=\mu_{BLM}^2\approx Q^2/9$. 
%Based on the same distribution, the complete NLO prediction
%for the pion transition form factor
%is calculated in the 
%$\alpha_V$ definition of the QCD coupling renormalized at 
%$\muR=\mu_V^2 = e^{5/3} \mu_{BLM}^2 \approx Q^2/2$.
%It is in good agreement with the presently available
%experimental data.
\end{abstract}

\section{Introduction}
\label{s:intro}

The pion transition form factor,
the simplest exclusive quantity, 
offers an excellent testing ground
for QCD. 
%
%It appears in the
%amplitude that relates two
%photons with the pion.
For large virtualities of the photons 
(or at least for one of them) perturbative QCD (PQCD) is
applicable \cite{LeBr80}.
A specific feature of this process
is that the leading-order (LO) prediction 
is zeroth order in the QCD coupling
constant, 
and one expects that PQCD for this
process may work at accessible values of
spacelike photon virtualities.

The pion transition form factor
$F_{\gamma \pi}(Q^2)$ is 
defined in terms
of the amplitude 
$\gamma^*(q,\mu) + \gamma(k,\nu) \rightarrow \pi(P)$
\begin{equation}
    \Gamma^{\mu \nu}=i \, e^2 \; F_{\gamma \pi}(Q^{2}) 
         \; \eps^{\mu \nu \alpha \beta} \; P_{\alpha}  q_{\beta}
               \, ,
\label{eq:Gmunu}
\end{equation}
and for large-momentum transfer $Q^2=-q^2$, 
it
can be represented \cite{excfw,LeBr80} 
as a convolution 
\begin{equation}
    F_{\gamma \pi}(Q^{2})= 
       \Phi^{*}(x,\muF) \, \otimes \, T_{H}(x,Q^{2},\muF) 
                    \,, 
\label{eq:tffcf}
\end{equation}
where $\otimes$ stands for the usual convolution symbol
%\begin{equation}
($A(z) \otimes B(z) = \int_0^1 dz A(z) B(z) \,$).
%\label{eq:defconv}
%\end{equation}
In \req{eq:tffcf}, the function
$T_{H}(x,Q^{2},\muF)$ is 
the hard-scattering amplitude
for producing a collinear $q \overline{q}$ pair
from the initial photon pair;
$\Phi^{*}(x,\muF)$ is the pion distribution amplitude
(DA) representing 
the probability amplitude for finding the valence
$q \overline{q}$ Fock state in the final pion with the 
constituents carrying fractions $x$ and $(1-x)$
of the meson's total momentum $P$; 
$\muF$ is
the factorization (or separation) scale at which
soft and hard physics factorize.
In this standard hard-scattering approach, 
pion is regarded as consisting only of valence Fock states, 
transverse quark momenta are neglected 
as well as quark masses.

The hard-scattering amplitude $T_{H}$ is obtained by
evaluating the  $\gamma^* \gamma \to q \overline{q}$
amplitude, 
and has a well--defined expansion in $\alpha_S(\muR)$,
with $\muR$ being the renormalization (or coupling constant) scale
of the hard-scattering amplitude.
Thus, one can write
\begin{eqnarray}
  T_{H}(x,Q^2,\muF)
       &=&
         T_{H}^{(0)}(x,Q^2)
         + \frac{\alpha_{S}(\muR)}{4 \pi} \, 
             T_{H}^{(1)}(x, Q^2,\muF) 
             \nonumber \\ & &
         + \frac{\alpha_{S}^2(\muR)}{(4 \pi)^2} \, 
             T_{H}^{(2)}(x, Q^2,\muF,\muR) 
                + \cdots  \, . 
\label{eq:TH}
\end{eqnarray}

The process-independent function $\Phi(x,\muF)$ is intrinsically 
nonperturbative, but
%(containing the effects of confinement, 
%nonperturbative interactions,
%and meson bound--state dynamics), 
it  satisfies an evolution equation of the
form
\begin{equation}
  \muF \frac{\partial}{\partial \muF} \Phi(x,\muF)   =
   V(x,u,\alpha_S(\muF)) \, \otimes \, \Phi(u,\muF)
         \, ,
\label{eq:eveq}
\end{equation}
where $V(x,u,\alpha_S(\muF))$ is the 
perturbatively calculable evolution kernel.
If the distribution amplitude $\Phi(x,\muO)$ 
is determined at an initial momentum scale $\muO$ 
(using some nonperturbative methods),
then the differential-integral evolution equation \req{eq:eveq} 
can be integrated using the moment method to give $\Phi(x,\muF)$.

The perturbative expansion of the pion transition
form factor takes the form
\begin{eqnarray}
F_{\gamma \pi}(Q^2)
      &=& 
       F_{\gamma \pi}^{(0)}(Q^2)
        + \frac{\alpha_S(\muR)}{4 \pi}  \, 
       F_{\gamma \pi}^{(1)}(Q^2)
        + \, \frac{\alpha_S^2(\muR)}{(4 \pi)^2} 
           \, 
       F_{\gamma \pi}^{(2)}(Q^2,\muR)
         + \cdots 
             \, .
\label{eq:Fgammapiexp}
\end{eqnarray}

The choice of the expansion parameter represents the
major ambiguity in the interpretation of the 
perturbative QCD predictions.
We see that the coupling constant $\alpha_S(\muR)$, as well as, 
the coefficients $F_{\gamma \pi}^{(i)}$ ($i>1$)
from \req{eq:Fgammapiexp}, 
depend on
the definition of the renormalization
scale and scheme.
%Naturally, all order result is independent
%of the scheme and scale choice but 
The truncation of the perturbative expansion 
at any finite order causes the residual dependence 
of the prediction on the choice of the 
renormalization scale and scheme, 
and introduces the theoretical uncertainty. 
If one can optimize the choices of the scale and scheme
according to some sensible criteria,
the size of the higher-order correction as well as the size of
the expansion parameter, i.e. the QCD running coupling
constant, can then serve as sensible indicators of the convergence
of the perturbative expansion. 

The simplest and widely used choice 
(the justification for the use of which is mainly pragmatic), 
is to take the
$\muR$ scale equal to characteristic momentum transfer 
of the process, i.e. in our case $\muR=Q^2$.
But since
each external momentum entering an exclusive reaction
is partitioned among many propagators of the
underlying hard-scattering amplitude, the physical scales
that control these processes are inevitably much softer
than the overall momentum transfer.

Several scale setting procedure were proposed in the literature
\cite{FAC,PMS,BLM83}.
In the Brodsky-Lepage-Mackenzie (BLM) 
procedure \cite{BLM83},
all vacuum-po\-la\-ri\-zat\-ion effects from the
QCD $\beta$-function are resummed 
into the running coupling constant. 
%Since the coefficients $\beta_0, \beta_1, \cdots $ are functions
%of $n_f$ (number of flavors),
According to BLM procedure,
the renormalization scale best suited
to a particular process in a given order can be, in practice,  
determined by computing vacuum-polarization insertions 
in the diagrams of that order, and by setting the scale
demanding that $n_f$-proportional terms should vanish.
The optimization of the renormalization scale and scheme
for exclusive processes by employing the BLM scale fixing
was elaborated  in \cite{BrJ98}.
The renormalization scales in the BLM method are physical
in the sense that they reflect the mean virtuality
of the gluon propagators 
and the important advantage of this method is
``pre-summing'' the large ($\beta_0 \alpha_S$)$^n$ terms,
i.e., the infrared renormalons associated with coupling constant
renormalization (\cite{BrJ98} and references therein). 

In our recent work \cite{MNP01} we have determined
the BLM scale for the pion transition form factor, 
i.e., for the $\gamma^* \gamma \rightarrow \pi$ process.
The LO prediction for the pion transition form factor 
is zeroth order in the QCD coupling constant, 
the NLO corrections \cite{tffNLO} 
represent leading QCD corrections 
and the vacuum polarization contributions
appearing at the next-to-next-to-leading order
(NNLO) were needed to fix the BLM scale from
from the requirement
\begin{equation}
F_{\gamma \pi}^{(2,n_f)}(Q^2,\muR=\mu_{BLM}^2)=0
     \, ,
\label{eq:BLMtff}
\end{equation}
where $F_{\gamma \pi}^{(2,n_f)}(Q^2,\muR)$
represents the $n_f$-proportional NNLO term from
\req{eq:Fgammapiexp}.  

In this work we outline important points of this
calculation and present the results that follow from the consistent
calculation up to $n_f$-proportional NNLO contributions
to both the hard-scattering and the distribution amplitude.

\section{Analytical calculation}
\label{s:cprocedure}

\begin{figure}[t]
  \centerline{\includegraphics[height=150pt]{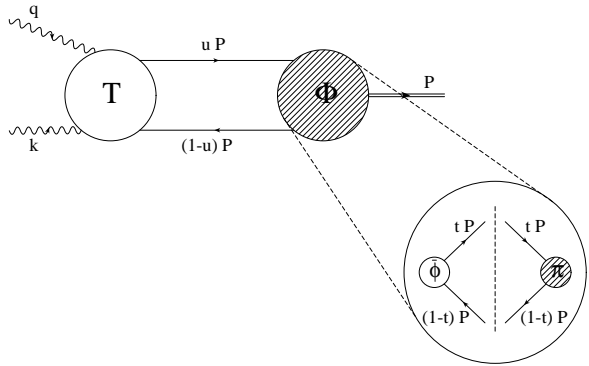}}
 \caption{Pictorial representation 
    of the pion transition form factor 
    calculational ingredients.}
 %   $T$ represents the perturbatively calculable hard-scattering
 %   amplitude , while $\Phi$ is the pion distribution
 %   amplitude given by \protect\req{eq:PhiOPi} which can be
 %   expressed, as in \protect\req{eq:Phitphirest},
 %   in terms of the perturbatively calculable part
 %   $\tilde{\phi}$
 %   \protect\req{eq:phiOqq}
 %   and the perturbatively uncalculable part.}  
 \label{f:FpiDA}
\end{figure}

We first outline the calculational procedure and
its ingredients which are illustrated 
in Fig. \ref{f:FpiDA}.

The 
$\gamma^* + \gamma \rightarrow q \overline{q}$
amplitude denoted by $T$
contains collinear singularities,
%(owing to the fact that final state quarks are taken to be
%massless and on-shell) 
and it factorizes as
\begin{equation}
    T(u,Q^2) = T_H(x, Q^2, \muF) \, \otimes \, Z_{T,col}(x, u; \muF)
        \, .
\label{eq:TTHZ}
\end{equation}
Here, 
$\muF$ denotes a factorization scale at which the separation
of collinear singularities takes place, and
all collinear singularities are factorized in $Z_{T,col}$,
since $T_H$ is, by definition, a finite quantity.

On the other hand, 
a process-independent distribution amplitude
for a pion 
in a frame where
$P^+=P^0+P^3=1$, $P^-=P^0-P^3=0$, and $P_{\perp}=0$
is defined \cite{LeBr80,Ka85etc} as
\begin{equation}
 \Phi (u)
  =   \int \frac{dz^-}{2 \pi} e^{i(u-(1-u))z^- /2}
    \, \left< 0 \left| 
  \bar{\Psi}(-z) \, \frac{\gamma^+ \gamma_5}{2\sqrt{2}}
           \, \Omega \, \Psi(z) 
    \right| \pi \right> _{(z^+=z_{\perp}=0)}
                \, , \qquad 
\label{eq:PhiOPi}
\end{equation}
where 
$
 \Omega  =  
  \textnormal{exp} \left\{ i g \int_{-1}^{1} ds A^+(z s)z^-/2 \right\}
$
is a path-ordered factor
making $\Phi$ gauge invariant.
The unrenormalized
pion distribution amplitude $\Phi(u)$ 
given in \req{eq:PhiOPi} and the distribution amplitude
$\Phi(v,\muF)$ renormalized at the scale 
$\muF$ are 
related by a multiplicative renormalizability equation
\begin{eqnarray}
   \Phi(u) & = &
            Z_{\phi,ren}(u,v; \muF) \otimes
              \Phi(v, \muF)
             \, .
\label{eq:PhiZPhi}
\end{eqnarray}

By convoluting
the ``unrenormalized'' (in the sense of collinear singularities)
hard-scattering amplitude $T(u,Q^2)$   
with the unrenormalized pion distribution amplitude
$\Phi(u)$,
given by \req{eq:TTHZ} and \req{eq:PhiZPhi}, respectively,
one obtains 
\begin{equation}
  F_{\gamma \pi}(Q^2) =
       \Phi^{\dagger}(u) \, \otimes \,  T(u, Q^2)
          \, .
\label{eq:Fpiur1}
\end{equation}
The divergences of $T(u,Q^2)$ and $\Phi(u)$ cancel
\begin{equation}
      Z_{T,col}(x,u; \muF) \otimes Z_{\phi, ren}(u,v; \muF)
           = \delta(x-v)
              \, ,
\label{eq:ZTZf}
\end{equation}
and the usual expression \req{eq:tffcf} emerges.
It is worth pointing out that
the scale $\muF$ representing the boundary between the low- and 
high-energy parts in
\req{eq:tffcf} 
is, at same time, the separation scale for
collinear singularities
in $T(u,Q^2)$, on the one hand,
and the renormalization scale for
UV singularities appearing in $\Phi(u)$, on the other hand.

We note also that the pion distribution amplitude as given in
\req{eq:PhiOPi}, with $\left| \pi \right>$ being the physical
pion state, of course, cannot be determined using perturbation theory. 
We can write $\Phi(u)$ as
\begin{equation}
   \Phi(u)  = 
       \tilde{\phi}(u,t) \otimes
            \, \left< q\bar{q}; t | \pi \right> \, 
               \, ,
\label{eq:Phitphirest}
\end{equation}
where $\tilde{\phi}(u,t)$ is obtained from \req{eq:PhiOPi}
by replacing the meson state $\left| \pi \right>$ 
by a $\left| q \overline{q}; t \right>$ state composed of a 
free quark and antiquark. 
The amplitude $\tilde{\phi}$  can be treated perturbatively,
making it possible to investigate
the high-energy tail of
the pion DA, to obtain $Z_{\phi,ren}$ 
and to determine the DA evolution.

We proceed to calculation.
This is the first calculation of the hard-scattering amplitude
$T(u,Q^2)$ of an exclusive process with the NNLO terms taken into account.
The subtraction (separation) of collinear divergences 
at the NNLO is significantly more demanding than that at the NLO.
Owing to the fact that the process under consideration
contains one pseudoscalar meson, the calculation is further
complicated by the $\gamma_5$ ambiguity related to the use
of the dimensional regularization method to treat 
UV and collinear divergences.
The consistent calculation of $T$ and $\tilde{\phi}$ enable us to
resolve these problems and, hence,
we have calculated
the LO, NLO, and $n_f$-proportional NNLO contributions to the 
perturbative expansions of both the hard-scattering
amplitude and the perturbatively calculable
part of the distribution amplitude.

\section{Discussing the factorization scale 
independence of the finite order result}

The dependence of pion distribution amplitude $\Phi(x,\muF)$
on $\mu_F^2$ is  specified 
by the evolution equation \req{eq:eveq}.
This dependence is completely contained in the evolutional
part $\phi_V$  
\begin{equation}
   \Phi(v,\muF)=
         \phi_V(v,s; \muF, \mu_0^2) \otimes
         \Phi(s, \mu_0^2)
           \, ,
\label{eq:PhifVPhi}
\end{equation}
which satisfies the evolutional equation 
\begin{equation}
       \muF
       \frac{\partial}{\partial \muF} 
                        \phi_V (v,s,\muF,\muO)
        = V(v,s',\muF) \, \otimes \, 
                           \phi_V(s',s,\muF,\muO)
             \, , \qquad 
\label{eq:VfV}
\end{equation}
while $\Phi(s, \mu_0^2)$ 
represents the nonperturbative input 
determined at the scale $\mu_0^2$. 

By differentiating \req{eq:tffcf} with respect to $\muF$ and
by taking into account \req{eq:eveq},
one finds that the hard-scattering amplitude satisfies
the evolution equation
\begin{equation}
  \muF \frac{\partial}{\partial \muF} 
        T_H(x, Q^2, \muF) = -
        T_H(y, Q^2, \muF)  \, \otimes \,
            V(y,x;\muF) 
         \, ,  
\label{eq:EvEqT}
\end{equation}
which is similar to \req{eq:eveq}.
The $\muF$ dependence of $T_H(x, Q^2, \muF)$
can be, analogous to \req{eq:PhifVPhi}, 
factorized in the function $\phi_V(y,x,Q^2,\muF)$ as
\begin{equation}
  T_H(x,Q^2,\muF) =  T_H(y,Q^2,\muF=Q^2) \otimes
                   \phi_V(y,x,Q^2,\muF)
            \, .
\label{eq:tTHfV}
\end{equation}
Using \req{eq:VfV} 
one can show by partial integration that
\req{eq:tTHfV} 
indeed represents the solution of
the evolution equation \req{eq:EvEqT}.

By substituting \req{eq:PhifVPhi} and \req{eq:tTHfV} 
in \req{eq:tffcf}, we obtain
\begin{equation}
  F_{\gamma \pi}(Q^2)
     =  
   T_H(y, Q^2, Q^2) 
          \, \otimes \phi_V(y,s, Q^2, \mu_0^2) 
          \, \otimes \,\Phi^*(s, \mu_0^2)
                 \, , \qquad 
\label{eq:Fpi2q2}
\end{equation}
where
\begin{equation}
   \phi_V(y,x, Q^2, \muF)\, \otimes \phi_V(x,s,\muF,\mu_0^2) 
          = \phi_V(y,s, Q^2, \mu_0^2) 
               \, ,
\label{eq:fvfvfv}
\end{equation}
has been taken into account. 
It is important to realize that the expression 
\req{eq:fvfvfv} is valid at 
every order of a PQCD calculation, and this can be easily
shown (see \cite{MNP01}).
Hence, the factorization scale $\muF$ 
disappears from the final prediction at every order in $\alpha_S$
and therefore does not introduce any theoretical uncertainty.
The crucial point is that both the resummation of
$(\alpha_S \ln(\muF/\mu_0^2))^n$ terms in $\Phi$ as well
as the resummation  of $(\alpha_S \ln(Q^2/\muF))^n$
terms in $T_H$, have to be performed using
\req{eq:PhifVPhi} and \req{eq:tTHfV} along with the results
from \req{eq:VfV}.
We note here that
by adopting the common choice $\mu_F^2 = Q^2$,
we avoid the need
for the resummation of the $(\alpha_S \ln(Q^2/\muF))^n$ terms in 
the hard-scattering part, making the calculation simpler. 

%The true expansion parameter left is 
%$\alpha_S(\muR)$, 
%with $\muR$ representing
%the renormalization scale of the complete
%perturbatively calculable part of the pion
%transition form factor \req{eq:Fpi2q2}, i.e., 
%of
%\begin{equation} 
%T_H(s, Q^2,\muO)=
% T_H(y, Q^2, Q^2) \otimes \phi_V(y,s, Q^2, \muO)
%    \, ,
%\label{eq:pertpart}
%\end{equation} 
%and we are left with
%the residual dependence on the $\muR$ scale, 
%when calculating to finite order.
 
\section{Numerical predictions}

We refer to \cite{MNP01} for the complete analytical
expressions for the pion transition form factor
calculated up to $n_f$ proportional NNLO terms.

\begin{figure}[t]
  \centerline{\includegraphics[height=5.5cm]{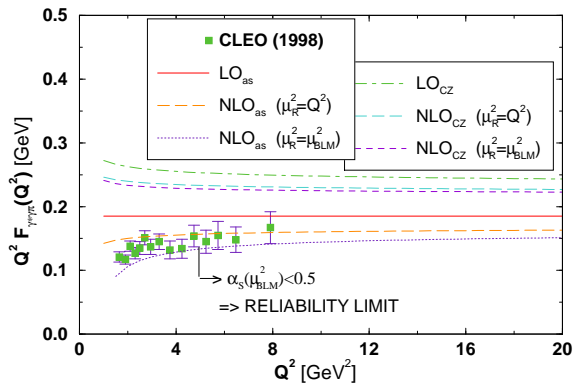}}
 \caption{The LO and NLO predictions for the pion transition
   form factor obtained using the $\overline{MS}$
   scheme  (and the usual one-loop  formula for $\alpha_S$
    with $\Lambda_{\overline{MS}}=0.2$ GeV$^2$).}
 \label{f:numASCZ}
\end{figure}
The prediction for the pion
transition form factor and the BLM scale $\mu_{BLM}^2$ 
depend on the form of the distribution amplitude.
There is increasing theoretical evidence coming from
different calculations
\cite{DA}
that the low energy pion distribution
amplitude does not differ much from its 
asymptotic form.
%$\phi_{as}(x) = 6 x (1-x)$.
%For comparison we also include the results
%obtained using the strongly
%end-point concentrated CZ distribution amplitude \cite{ChZ84}
%$\phi_{CZ}(x,\mu_0^2) = 6 x (1-x) \; [5 \, (2 x-1)^2]$
%($\mu_0^2=0.5$ GeV$^2$).

The expression for the pion transition form factor 
$Q^2 F_{\gamma \pi}(Q^2)$
corresponding to the asymptotic distribution
reads
\begin{eqnarray}
  Q^2 F_{\gamma \pi}(Q^2) 
     &=& 2 C_{\pi} f_{\pi}
     \left\{ 3  
    +\frac{\alpha_S(\muR)}{4 \pi}
       (-20) 
    +\frac{\alpha_S^2(\muR)}{(4 \pi)^2}
         \right. \nonumber \\ & & \left. \times
       \left[ \left( -\frac{2}{3} n_f \right)
       \left( -43.47 - 20 \ln \frac{\muR}{Q^2} \right)
         + \cdots \right] 
       + \cdots \right\}
            \, , \quad
\label{eq:numresAS}
\end{eqnarray}
where $C_{\pi}=\sqrt{2}/6$ is a flavour factor, while
$f_{\pi}=0.131$ GeV. 
The $n_f$-proportional NNLO contribution
determines the value of the BLM scale 
\begin{equation}
 \muR = \left(\mu_{BLM}^2\right)^{as}
   \approx \frac{Q^2}{9}
        \, .
\label{eq:BLMas}
\end{equation}
One notes that this scale is considerably softer
than the total momentum transfer $Q^2$, 
which is consistent with partitioning of $Q^2$
among the pion constituents.
%It should be pointed out, however, 
%that in the $\overline{MS}$ scheme
%the BLM scale does not correspond to the mean gluon momenta.
%Based on \req{eq:numresAS}, the NLO prediction amounts to
%\begin{equation}
%  Q^2 F_{\gamma \pi}(Q^2)= 
%      0.185 
%    +\frac{\alpha_S(\muR)}{\pi} (-0.309)
%    + \cdots
%        \, .
%\label{eq:numResAS}
%\end{equation}

The NLO predictions obtained 
in the $\overline{MS}$ scheme are displayed
in Fig. \ref{f:numASCZ}.
The predictions based on the asymptotic DA are,
in contrast to the  ones obtained using the CZ DA
\cite{ChZ84}, in good agreement with the 
experimental data \cite{Gr98ea}.

Nevertheless,
the rather low BLM scale given in \req{eq:BLMas},
and consequently the large $\alpha_S(\mu_{BLM}^2)$,
questions the applicability 
of the perturbative prediction at experimentally
accessible momentum transfers.
The NLO predictions obtained 
assuming the asymptotic DA and the BLM scale \req{eq:BLMas}
satisfy the requirement $\alpha_S(\muR)<0.5$
for $Q^2 \geq 6$ GeV$^2$.
This reliability limit is indicated on Fig. \ref{f:numASCZ}.
%It should be pointed out, however, that
%there is an intrinsic disadvantage
%in using the usual one-loop $\overline{MS}$ running coupling
%as an expansion parameter,
%since it has a simple pole at $\muR=\Lambda^2$ and 
%this does not reflect the nonperturbative
%behavior 
%of $\alpha_S(\muR)$ for small $\muR$. 
The transition to the more physical $\alpha_V$ scheme,
may offer a way out of this problem.

\begin{figure}[t]
  \centerline{\includegraphics[height=5.5cm]{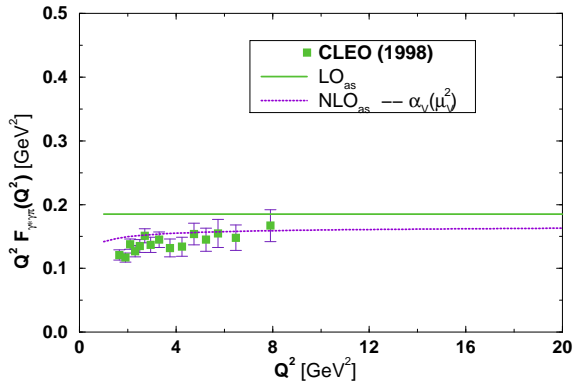}}
 \caption{The LO and NLO predictions for the pion transition
   form factor in the $\alpha_V$ scheme
   ($\Lambda_V=0.16$ GeV$^2$).}
 \label{f:numV}
\end{figure}
%
%So far nothing has been said concerning the
%renormalization scheme dependence of the predictions.
%It is known that the renormalization scheme dependence can be 
%avoided by considering relations between physical observables, 
%which must be independent of the choice of the scheme and scale
%to any fixed order of perturbation theory.
%This requirement can be expressed in the form of 
%"commensurate scale relations" (CSR), 
%in which the BLM scale-setting method is used to fix the 
%renormalization scale \cite{BrLu95}. 
%In practice, a CSR for two observables is obtained 
%by relating their respective perturbative predictions calculated in, 
%for example, the $\overline{MS}$ scheme, and then by algebraically 
%eliminating $\alpha_{\overline{MS}}$.
%The choice of the BLM scale ensures
%that the resulting CSR is independent of the choice 
%of the intermediate renormalization scheme.
%Following this approach, in \cite{BrJ98}
In \cite{BrJ98}
the exclusive hadronic amplitudes
were analysed in the
$\alpha_V$ scheme, 
in which the effective coupling $\alpha_V(\mu^2)$ is defined
from the heavy-quark potential $V(\mu^2)$.
The $\alpha_V$ scheme is a natural, physically based scheme, 
which by definition automatically incorporates vacuum polarization 
effects %due to the fermion pairs 
into the coupling. 
The $\mu_V^2$ scale %, 
%which then appears in the coupling, 
reflects the mean virtuality of the 
exchanged gluons.

If use is made of the scale-fixed relation 
between  the couplings 
$\alpha_{\overline{MS}}$ 
and $\alpha_V$ \cite{BrJ98}
%\begin{equation}
%\alpha_{\overline{MS}}(e^{-5/3} \mu_V^2)
%=
%\alpha_V(\mu_V^2) \left( 1 + \frac{\alpha_V(\mu_V^2)}{4 \pi}
%          \, \frac{8 C_A}{3} + \cdots \right) \, ,
%\label{eq:defalphaV}
%\end{equation}
then,
to the order we are calculating, 
the NLO prediction 
in the $\alpha_V$ scheme is obtained by taking 
$\muR= \mu_V^2 = e^{5/3} \, \mu_{BLM}^2$, i.e.
for the asymptotic DA
\begin{equation}
         (\mu_V^2)^{as} = e^{5/3} \, (\mu_{BLM}^2)^{as} 
          \approx \frac{Q^2}{2}
              \, .
\label{eq:Vsc}
\end{equation}
The NLO prediction for $Q^2 F_{\gamma \pi}(Q^2)$ obtained in
$\alpha_V$ scheme
is depicted in Fig. \ref{f:numV}.
As can be seen, it is in good agreement with experimental data.
We note that, since $\alpha_V$ is an effective running coupling defined
from the physical observable, it must be finite at low momenta,
and the appropriate parameterization of the low-energy region 
should in principle be included
(see \cite{alphaSmod} for various proposals).

%Nevertheless, in the energy region we are interested
%in, the usual one-loop conventional solution of the
%renormalization group equation for the QCD coupling 
%can be employed.
%The LO QCD correction, i.e., the NLO contribution,
%lowers the LO prediction for $\approx 16\%$ 
%for $Q^2 \approx 6$ GeV$^2$, i.e., for $\alpha_V(\mu_V^2) \approx 0.3$.

\section{Conclusions}
\label{s:concl}

In this paper we have reviewed the determination of 
the NLO BLM scale
for the pion transition form factor.
A consistent calculation 
of both the hard-scattering and the 
perturbatively calculable part of 
the pion distribution amplitude has been performed
up to $n_f$-proportional NNLO terms.

It has been demonstrated that the prediction 
for the pion transition form factor is 
independent of the factorization scale $\muF$ 
at every order in the strong coupling constant
$\alpha_S$. Provided both the hard-scattering and the
distribution amplitude are treated consistently 
regarding their $\muF$ dependence,
the factorization scale  
disappears from the final prediction at every order in $\alpha_S$
without introducing any theoretical uncertainty.
%Consequently, for practical purposes the
%simplest and commonly used choice $\muF=Q^2$
%is justified at the intermediate steps of the calculation.
One can use $\muF=Q^2$ to simplify the calculation,
but any other choice would lead to the same result.

%The NLO prediction derived from the asymptotic 
%distribution is in good agreement with the presently
%available experimental data, while the prediction
%obtained assuming the CZ distribution exceeds the data
%significantly, clearly demonstrating the inadequacy of the CZ
%distribution.

The renormalization scale $\muR$  fixed
according to the BLM scale setting prescription
within the $\overline{MS}$ scheme
and corresponding to the asymptotic pion distribution amplitude,
turns out to be
$\mu_{BLM}^2 \approx Q^2/9$.
Thus, in the region of $Q^2 < 8$ GeV$^2$,
in which the experimental data exist,
$\mu_{BLM}^2 < 1$ GeV$^2$.
Consequently, the prediction obtained with $\muR=\mu_{BLM}^2$
cannot, in this region, be considered reliable.

In addition to the results calculated 
in the $\overline{MS}$ renormalization
scheme, the numerical prediction assuming the same distribution
but in the  $\alpha_V$ scheme, 
with the renormalization scale 
$\muR=\mu_V^2 = e^{5/3} \mu_{BLM}^2 \approx Q^2/2$,
has also been obtained.
It is displayed in Fig. \ref{f:numV} and,
as seen, is in good agreement with experimental data.
Due to the fact that the scale $\mu_V^2$ reflects
the mean gluon momentum in the NLO diagrams,
it is to be expected that the higher-order QCD
corrections are minimized, so that the leading order
QCD term gives a good approximation to the complete sum. \\[0.1cm]
\noindent{\bf Acknowledgments }
  One of us (B.M.) acknowledges the support 
  by the Alexander von Humboldt Foundation. 
  This work was supported by the Ministry of Science and Technology
  of the Republic of Croatia under Contract No. 0098002.

%INDEX%%%%%%%%%%%%%%%%%%%%%%%%%%%%%%%%%%%%%%%%%%%%%%%%%%%%%%%%%%%%%%%
% Please check with the editor of your book whether he plans to
% include a "mutual" subject index - if so, please code your entries
% in the standard syntax. For your own purposes you may print your
% "personal" index by using the following commands:
%
%\clearpage
%\addcontentsline{toc}{section}{Index}
%\flushbottom
%\printindex
%%%%%%%%%%%%%%%%%%%%%%%%%%%%%%%%%%%%%%%%%%%%%%%%%%%%%%%%%%%%%%%%%%%%%


\begin{thebibliography}{8.}

\bibitem{LeBr80}
G.~P.~Lepage and S.~J.~Brodsky,
%``Exclusive Processes In Perturbative Quantum Chromodynamics,''
Phys.\ Rev.\ D {\bf 22}, 2157 (1980)
%%CITATION = PHRVA,D22,2157;%%

\bibitem{excfw}
G.~P.~Lepage and S.~J.~Brodsky,
%``Exclusive Processes In Quantum Chromodynamics: Evolution Equations 
% For Hadronic Wave Functions And The Form-Factors Of Mesons,''
Phys.\ Lett.\ B {\bf 87}, 359 (1979);
%%CITATION = PHLTA,B87,359;%%
%
A.~V.~Efremov and A.~V.~Radyushkin,
%``Factorization And Asymptotical Behavior Of Pion Form-Factor In QCD,''
Phys.\ Lett.\ B {\bf 94}, 245 (1980);
%%CITATION = PHLTA,B94,245;%%
%
A.~Duncan and A.~H.~Mueller,
%``Asymptotic Behavior Of Exclusive And Almost Exclusive Processes,''
Phys.\ Lett.\ B {\bf 90}, 159 (1980)
%%CITATION = PHLTA,B90,159;%%
%

\bibitem{FAC}
G.~Grunberg,
%``Renormalization Scheme Independent QCD And QED: The Method Of
%Effective Charges,''
Phys.\ Rev.\ D {\bf 29}, 2315 (1984)
%%CITATION = PHRVA,D29,2315;%%

\bibitem{PMS}
P.~M.~Stevenson,
%``Optimization And The Ultimate Convergence Of QCD Perturbation
%Theory,''
Nucl.\ Phys.\ B {\bf 231}, 65 (1984)
%%CITATION = NUPHA,B231,65;%%

\bibitem{BLM83}
%\bibitem{BrL83}
S.~J.~Brodsky, G.~P.~Lepage and P.~B.~Mackenzie,
%``On The Elimination Of Scale Ambiguities In 
%  Perturbative Quantum Chromodynamics,''
Phys.\ Rev.\ D {\bf 28}, 228 (1983)
%%CITATION = PHRVA,D28,228;%%

\bibitem{BrJ98}
%\bibitem{BroJ98}
S.~J.~Brodsky, C.~Ji, A.~Pang and D.~G.~Robertson,
%``Optimal renormalization scale and scheme for exclusive processes,''
Phys.\ Rev.\ D {\bf 57}, 245 (1998)
% [hep-ph/9705221].
%%CITATION = HEP-PH 9705221;%%

\bibitem{MNP01}
B.~Meli\'{c}, B.~Ni\v{z}i\'{c} and K.~Passek,
%``BLM scale for the pion transition form factor,''
Phys.\ Rev.\ D {\bf 65}, 053020 (2002);
%[arXiv:hep-ph/0107295].
%%CITATION = HEP-PH 0107295;%%
%
%B.~Meli\'{c}, B.~Ni\v{z}i\'{c} and K.~Passek,
%``A note on the factorization scale independence of the PQCD
%predictions  for exclusive processes,''
hep-ph/0107311
%%CITATION = HEP-PH 0107311;%%

\bibitem{tffNLO}
F.~del Aguila and M.~K.~Chase,
%``Higher Order QCD Corrections To Exclusive Two Photon Processes,''
Nucl.\ Phys.\  {\bf B193}, 517 (1981);
%%CITATION = NUPHA,B193,517;%%
%
E.~Braaten,
%``QCD Corrections To Meson - Photon Transition Form-Factors,''
Phys.\ Rev.\ D {\bf 28}, 524 (1983);
%%CITATION = PHRVA,D28,524;%%
%
E.~P.~Kadantseva, S.~V.~Mikhailov and A.~V.~Radyushkin,
%``Total Alpha-S Corrections To Processes Gamma* Gamma* $\to$ Pi0 And Gamma* Pi $\to$ Pi In A Perturbative QCD,''
Yad.\ Fiz.\  {\bf 44}, 507 (1986)
[Sov.\ J.\ Nucl.\ Phys.\  {\bf 44}, 326 (1986)]
%%CITATION = YAFIA,44,507;%%

\bibitem{Ka85etc}
G.~R.~Katz,
%``Two Loop Feynman Gauge Calculation Of The Meson Nonsinglet Evolution Potential,''
Phys.\ Rev.\ D {\bf 31}, 652 (1985);
%%CITATION = PHRVA,D31,652;%%
%
S.~J.~Brodsky, P.~Damgaard, Y.~Frishman and G.~P.~Lepage,
%``Conformal Symmetry: Exclusive Processes Beyond Leading Order,''
Phys.\ Rev.\ D {\bf 33}, 1881 (1986)
%%CITATION = PHRVA,D33,1881;%%

\bibitem{DA}
V.~Braun and I.~Halperin,
%``Soft contribution to the pion form-factor from light cone QCD sum rules,''
Phys.\ Lett.\ B {\bf 328}, 457 (1994);
% [hep-ph/9402270].
%%CITATION = HEP-PH 9402270;%%
%
R.~Jakob, P.~Kroll and M.~Raulfs,
%``Meson - photon transition form-factors,''
J.\ Phys.\ G {\bf 22}, 45 (1996);
% [hep-ph/9410304].
%%CITATION = HEP-PH 9410304;%%
%
A.~V.~Radyushkin,
%``{QCD} sum rules and soft-hard interplay for hadronic form
%factors,''
Few Body Syst.\ Suppl.\  {\bf 11}, 57 (1999);
%[arXiv:hep-ph/9811225].
%%CITATION = HEP-PH 9811225;%%
%
A.~Schmedding and O.~Yakovlev,
%``Perturbative effects in the form factor gamma gamma* $\to$ pi0 
% and  extraction of the pion wave function from CLEO data,''
Phys.\ Rev.\ D {\bf 62}, 116002 (2000);
% [hep-ph/9905392].
%%CITATION = HEP-PH 9905392;%%
%
A.~P.~Bakulev, S.~V.~Mikhailov and N.~G.~Stefanis,
%``QCD-based pion distribution amplitudes confronting experimental data,''
Phys.\ Lett.\ B {\bf 508}, 279 (2001)
% [hep-ph/0103119].
%%CITATION = HEP-PH 0103119;%%

\bibitem{ChZ84}
V.~L.~Chernyak and A.~R.~Zhitnitsky,
%``Asymptotic Behavior Of Exclusive Processes In QCD,''
Phys.\ Rept.\  {\bf 112}, 173 (1984)
%%CITATION = PRPLC,112,173;%%

\bibitem{Gr98ea}
J.~Gronberg {\it et al.}  [CLEO Collaboration],
%``Measurements of the meson photon transition form factors of light  
% pseudoscalar mesons at large momentum transfer,''
Phys.\ Rev.\ D {\bf 57}, 33 (1998)
% [hep-ex/9707031].
%%CITATION = HEP-EX 9707031;%%

\bibitem{alphaSmod}
J.~M.~Cornwall,
%``Dynamical Mass Generation In Continuum QCD,''
Phys.\ Rev.\ D {\bf 26}, 1453 (1982);
%%CITATION = PHRVA,D26,1453;%%
%
A.~Donnachie and P.~V.~Landshoff,
%``Gluon Condensate And Pomeron Structure,''
Nucl.\ Phys.\  {\bf B311}, 509 (1989);
%%CITATION = NUPHA,B311,509;%%
D.~V.~Shirkov and I.~L.~Solovtsov,
%``Analytic model for the QCD running coupling with universal  
% alpha(s)-bar(0) value,''
Phys.\ Rev.\ Lett.\  {\bf 79}, 1209 (1997)
% [hep-ph/9704333].
%%CITATION = HEP-PH 9704333;%%
%

\end{thebibliography}
\end{document}